# Triple Ladder Lumped Circuit with Sixth Order Modal Exceptional Degeneracy

Farshad Yazdi, Alireza Nikzamir, Tarek Mealy, Mohamed Y. Nada, and Filippo Capolino

***Abstract*—We introduce a circuit topology based on a simple triple-ladder circuit realized with lumped reactive components that provides a sixth order degenerate band-edge (6DBE). The 6DBE is a special kind of sixth-order exceptional point of degeneracy in a lossless and gainless periodic ladder. This degeneracy provides a very flat band edge in the phase-frequency dispersion diagram. The proposed topology exhibits unique structured resonance features associated with a high loaded Q-factor. We investigate the Floquet-Bloch modes in an infinite-length periodic triple-ladder and their dispersion relation using the S parameter formalism. We also provide the approximate analytic expressions of the eigenmodes and dispersion relation around the degenerate point based on the Puiseux series expansion. We investigate the filtering characteristics of a finite-length structure terminated with loads to highlight the special properties of the 6DBE compared to ladders with regular band edge (RBE) and fourth order degenerate band edge (DBE). The circuit framework introduced here with a 6DBE can be exploited in designing novel high Q-factor oscillators, filters, sensors, and pulse shaping networks.**

***Index Terms*—Sixth order degeneracy, 6DBE, triple-ladder circuit, lumped circuit.**

## I. Introduction

We show the theory of a periodic circuit that exhibits a sixth-order degenerate band edge (6DBE) in the dispersion relation and explore the associated unique characteristics. The 6DBE is a point in the system parameter space at which six eigenmodes coalesce in both the eigenvalues and eigenvectors. We demonstrate a simple realization and design procedure to have a 6DBE, which can be easily applied to other technologies such as periodic microstrip transmission lines. Degeneracy in electromagnetic waveguides means that independent eigenmodes coalesce at a certain angular frequency, denoted by $\omega_e$, and form a single degenerate mode. Mathematically, this special degeneracy is manifested when the system matrix describing propagation is defective and a complete basis of eigenvectors cannot be obtained [1], [2].

A special and well-known category of degeneracies in periodic structures is recognized as the fourth order degenerate band edge (DBE) where four periodic eigenstates coincide at the center or edge of the Brillouin zone [3]–[7]. The existence and characteristics of the fourth order DBE have been exhibited in various periodic structures including transmission lines and microstrip lines [7]–[9], as well as metallic [10] and optical waveguides [11]–[14]. Experimental demonstrations of the DBE at microwave frequencies are in [7], [10]. In [15], a double ladder periodic lumped circuit is introduced to exhibit a DBE, explaining the advantages of high quality factor and stability to load changes. In this paper, we introduce a periodic lumped circuit based on a triple-ladder design that develops a sixth order degeneracy. This condition happens when six Floquet-Bloch modes (eigenstates) coalesce at the edge of the Brillouin zone. At this condition, the dispersion relation around the degeneracy point is approximated by $(\omega_e - \omega) \propto (\varphi - \pi)^6$ in which $\varphi$ is the shift in the phase from one unit cell of the periodic structure to its adjacent one, $\omega$ is the angular frequency, and $\omega_e$ is the frequency that features the sixth-order degeneracy. In [16], a scheme for a novel oscillator design based on the DBE was shown to lead to a lower oscillation threshold and robustness to the effects of loss and loading as opposed to ladders featuring only the regular band edge [6], [17] based on the enhanced quality factor associated to the DBE condition. Remarkable characteristics of the degeneracy condition in the periodic structure including the significant decrease in the group velocity and sharp increase of the quality factor and local density of states [6], [16], will make this class of cavities with degeneracy suitable for a variety of applications ranging from sensors and filters to oscillator and pulse shaping circuits [18]–[22].

The occurrence of the 6th order degeneracy is demonstrated in an optical waveguide made of couple ring optical resonators in [23], via simulations and coupled mode theory. In this paper, we focus on a periodic circuit design with lumped elements which can support a 6DBE in its phase-frequency dispersion relation. We also focus on some important properties related to the loaded quality factor as discussed next. The unit cell design of the periodic structure proposed here comprises a six-port network composed of reactive components (capacitors and inductors) and has coupling between the branches, as shown in Fig. 1(a). Due to the simplicity of the proposed unit cell, we are able to offer the analytic formulation of the eigenvalues (hence the modes) based on the Puiseux fractional power series expansion around the degeneracy condition [24]. The general framework developed here is not limited to one specific design or application and can be potentially applied to various configurations featuring degenerate band edge phenomena.

In Section II, we introduce the triple-ladder circuit and provide a detailed investigation of the modes and dispersion relation around the degeneracy condition for the infinitely long periodic circuit. An approximation of the system's eigenvalues based on the Puiseux series expansion is also provided. In Section III, we investigate the finite-length structure's properties including the transfer function of the system and its resonance behavior around the 6DBE. We also account for the effects of losses as well as $Q$-factor scaling with the length of the resonator, including loss/loading effects, analogously to what was done for the 4th order DBE in [15].

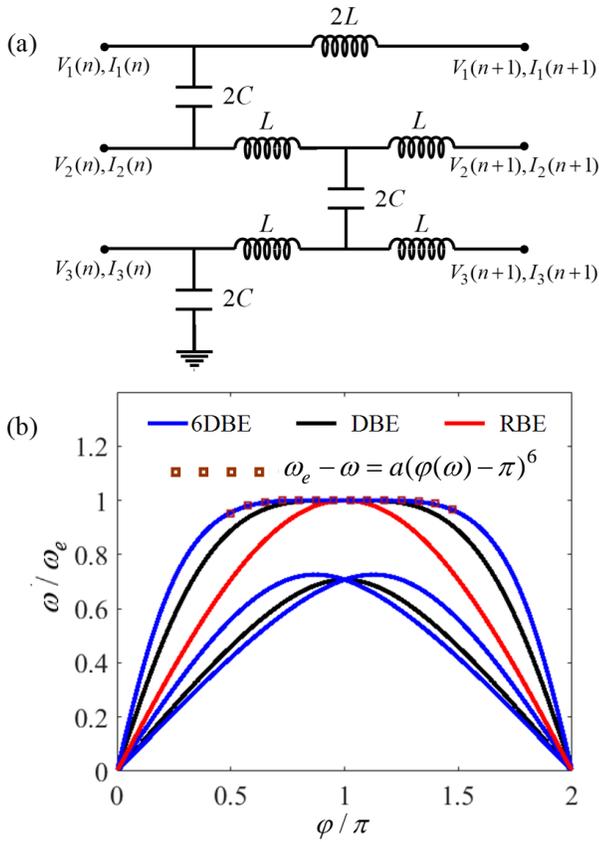

Fig. 1. (a) Unit cell of a periodic triple-ladder lumped circuit that develops a sixth order degeneracy (6DBE) at an angular frequency $\omega_e = 1/\sqrt{LC}$. (b) Dispersion diagram of the six eigenmodes in the infinitely long periodic triple-ladder circuit exhibiting a 6DBE at $\omega_e$. The fitting curve (red squares) shows analytically the flatness of the dispersion around the 6DBE. The 6DBE is flatter than the DBE and the RBE.

## II. TRIPLE LADDER CIRCUIT

The proposed design for the unit cell of a periodic circuit that can support a sixth order degeneracy is presented in Fig. 1(a), composed of eight reactive components (five inductors and three capacitors) that comprises a six-port network. As shown in Fig. 1(a), the coupling between the three ladders is provided by the capacitors and this design will develop a 6DBE at angular frequency $\omega_e = 1/\sqrt{LC}$. After designing the unit cell featuring 6DBE at the desired frequency with practical values for inductors and capacitors, we will study the modal characteristics and dispersion relation of the triple ladder circuit in the following subsection, where we also provide the mathematical analysis of the system's behavior in the vicinity of the degeneracy condition. For all the numerical results provided in this paper, we are assuming $C = 56$ pF and $L = 45$ nH in our unit cell design featuring a 6DBE at the frequency $f_e = 100.26$ MHz.

### A. State Vector and Dispersion Relation

To investigate the modal characteristics (namely, eigenvalues and eigenvectors) of the periodic structure, we exploit the transfer matrix formalism which relates the voltages and current from one unit cell to the next. In order to do so, we have defined the state vector $\Psi(n)$ comprising of the voltages and current phasors at the three input ports of the arbitrary $n$-th unit cell shown in Fig. 1(a) as

$$\Psi(n) = [V_1(n), \; V_2(n), \; V_3(n), \; I_1(n), \; I_2(n), \; I_3(n)]^T. \quad (1)$$

The state vector is translated across the unit cell by the 6×6 forward transfer matrix $\underline{\mathbf{T}}$ as

$$\Psi(n+1) = \underline{\mathbf{T}} \Psi(n). \quad (2)$$

The detailed calculation of the transfer matrix for the triple ladder circuit proposed in Fig. 1(a) is provided in the Appendix.

As discussed in [15] for infinitely long periodic circuits, the periodic solutions for the eigenstates $\Psi(n)$ at location $n$ (or at $n$-th unit cell) transform to the state vector $\Psi(n+1)$ at location $n+1$ (or at $(n+1)$-th unit cell), satisfy $\Psi(n+1) = e^{-j\varphi(\omega)}\Psi(n)$, where $\varphi(\omega)$ for a lossless structure is simply the Floquet-Bloch phase shift between two consecutive unit cells [15]. In general, for the triple ladder network we get six different solutions for $\varphi(\omega)$ at every frequency, each of which represents a Floquet-Bloch mode of the system. Those periodic solutions and hence the dispersion relation (between the phase shift and the frequency) are obtained by solving the eigenvalue equation

$$\left[\underline{\mathbf{T}} - e^{-j\varphi(\omega)}\underline{\mathbf{I}}\right]\Psi(n) = 0, \quad (3)$$

in which $\underline{\mathbf{I}}$ is the 6×6 identity matrix. The dispersion diagram of the circuit made of an infinite number of unit cells, shown in Fig. 1(a), is plotted in Fig. 1(b). This plot shows only the eigenstates with purely real phase shift $\varphi(\omega)$ normalized to $\pi$ versus normalized angular frequency. The proposed unit cell exhibits a 6DBE at the angular frequency $\omega_e = 1/\sqrt{LC}$ and phase shift of $\varphi(\omega_e) = \pi$. The 6DBE is observed by the flatness of the blue lines in Fig. 1(b); six Floquet-Bloch modes (eigenstates) coalesce at the edge of the Brillouin zone. To see the full coalescence at $\varphi(\omega_e) = \pi$ we would need to plot also the branches with complex $\varphi(\omega)$ that are omitted in this figure for simplicity. Around the 6DBE condition, the dispersion relation is well approximated by $(\omega_e - \omega) \propto (\varphi - \pi)^6$ as shown in the results of Fig. 1(b) as expected by the order of the degeneracy. To better perceive the flatness associated to the 6DBE condition in comparison to lower orders of degeneracy, the dispersion diagrams for the circuits exhibiting DBE and RBE are also depicted in Fig. 1(b). There, we have considered the double-ladder and single-ladder circuit designs provided in [15], and for them, we use the same component values as in our 6DBE design to get DBE and RBE, respectively.

### B. Analysis of the System Behavior in the Vicinity of the 6DBE

We provide an analytical approach to approximate the perturbed eigenvalues of the system and hence the dispersion relation when moving away from the ideal degeneracy condition due to the frequency perturbation, although the analytical framework provided here can be applied to other types of perturbations such as losses introduced to the circuit's elements as well. For the sake of our frequency perturbation analysis, we consider a lossless circuit and assume a change of the frequency of operation slightly away from the 6DBE angular frequency $\omega_e$, described by $\delta_\omega = (\omega - \omega_e)/\omega_e$. The eigenstates of the system (eigenvalues $\lambda = \exp(-j\varphi)$ of the transfer matrix $\mathbf{T}$) are then be approximated using the Puiseux fractional power series expansion around the ideal (unperturbed) degenerate eigenvalue $\lambda_e = -1$, following the equation [24], [25]

$$\lambda_p(\delta_\omega) = \lambda_e + \sum_{k=1}^{\infty} \alpha_k \left(\zeta^p \delta_\omega^{1/6}\right)^k, \quad (4)$$

where, $p = 1, 2, \ldots, 6$ and $\zeta = \exp(j\pi/3)$ provide six different solutions for the perturbed eigenvalues. The Puiseux series coefficients, $\alpha_k$, are calculated using the general recursive formulas provided in [24]. The first two series coefficients are obtained using the following formulas

$$\alpha_1 = \left(-\frac{\frac{\partial D}{\partial \delta_\omega}(0, \lambda_e)}{\frac{1}{6!}\frac{\partial^6 D}{\partial \lambda^6}(0, \lambda_e)}\right)^{1/6} \neq 0,$$

$$\alpha_2 = \frac{-\left(\alpha_1^7 \frac{1}{7!}\frac{\partial^7 D}{\partial \lambda^7}(0, \lambda_e) + \alpha_1 \frac{\partial^2 D}{\partial \lambda \partial \delta_\omega}(0, \lambda_e)\right)}{6\alpha_1^5 \left(\frac{1}{6!}\frac{\partial^6 D}{\partial \lambda^6}(0, \lambda_e)\right)}, \quad (5)$$

where $D(\delta_\omega, \lambda)$ is the characteristic polynomial of the transfer matrix $\mathbf{T}$ defined as

$$D(\delta_\omega, \lambda) = \det(\mathbf{T}(\delta_\omega) - \lambda \mathbf{I}); \quad D(\delta_\omega, \lambda) = \sum_{l=0}^{6} b_l \lambda^l. \quad (6)$$

This is the sixth order polynomial of $\lambda$ and the coefficients $b_l$ are calculated based on the applications of the Cayley-Hamilton theorem [26] stating that the matrices satisfy the characteristic equation. Due to the reciprocity of the system, if $\lambda_i$ is a solution, then $1/\lambda_i$ must also be a solution. Exactly at the 6DBE condition, all the eigenvalues $\lambda_i$ are equal, i.e., $\lambda_i = \lambda_e$, hence the only possible solution is either $\lambda_e = 1$ (i.e., at the edge of the Brillouin zone defined as $[0 \; 2\pi]$) or $\lambda_e = -1$ (i.e., at the center of the Brillouin zone). Considering the 6DBE at the center, i.e., $\lambda_e = -1$, the first two Puiseux series coefficients for the periodic triple-ladder circuit design of the unit cell shown in Fig. 1(a) are calculated and heavily simplified in terms of $L$ and $C$ as

$$\alpha_1 = \sqrt[6]{128}, \quad \alpha_2 = -\sqrt[3]{16}, \quad (7)$$

which by substituting the values of these coefficients in Eq. (4), the six eigenvalues around the 6DBE are approximated as

$$\lambda_p(\delta_\omega) \approx \lambda_e + \alpha_1 \left(\zeta^p \delta_\omega^{\frac{1}{6}}\right) + \alpha_2 \left(\zeta^p \delta_\omega^{\frac{1}{6}}\right)^2, \quad (8)$$

where here we only consider the first two terms of the expansion around $\lambda_e = -1$. (The Puiseux approximation used in the numerical results is based only on the first order, we provide also $\alpha_2$ because it has a very simple expression though in the examples provides there was no need to include it because the first term). This is a good approximation of the perturbed eigenvalues around the 6DBE, though more terms could be added to achieve even better approximations. By considering that $\lambda = \exp(-j\varphi)$, the dispersion relation $(\varphi, \omega)$ of the circuit can be approximated with the Puiseux series in terms of fractional powers of frequency perturbation. By considering only the first coefficient of the Puiseux series, the dispersion relation is approximated as $(\varphi - \pi) \approx \alpha_1 (\omega - \omega_e)^{1/6}$. By rearranging the derived dispersion relation, we get

$$(1 - \omega/\omega_e) \approx \eta (1 - \varphi/\pi)^6, \quad (9)$$

in which $\eta$ is a dimensionless parameter named "flatness factor" associated to the sixth derivative of $\omega$ with respect to $\varphi$ at $\omega_e$, indicating the flatness of the dispersion diagram at the exceptional point of degeneracy. For the circuit proposed in this paper, the flatness factor $\eta$ for the normalized dispersion relation is calculated to be equal to $\pi^6/128$. It is also worth noting that $\eta$ does not depend on the values of the circuit parameters, namely $L$ and $C$, thus changing the values of $L$ and $C$ will result in the same "normalized" dispersion relation and flatness factor around $\omega_e$. In other words, these results can be scaled to other 6DBE frequencies without altering the flatness of the dispersion. Figure 2 shows the complex dispersion diagram with both the real and the imaginary branches of $\varphi(\omega)$, assuming an infinitely-long periodic triple-ladder circuit comprised of the unit cell shown in Fig. 1(a). The red curves represent the approximated dispersion diagram derived via the first-order Puiseux series approximation of the eigenvalues, while the blue curves represent the "exact" eigenvalue, numerically calculated from the dispersion relation. The dispersion relations of Fig. 2 show a very good agreement between the numerical results and the analytical Puiseux series approximation around the sixth order degeneracy condition by using only the first term.

## III. RESONANCE BEHAVIOR OF THE CIRCUIT WITH FINITE NUMBER OF UNIT CELLS

To investigate the resonance behavior of the circuit with sixth order modal degeneracy of Fig. 1(a), we consider a finite-length structure made by cascading $N$ unit cells as shown in Fig. 3. We consider the triple-ladder structure with proper excitations and terminations to analyze some of its important characteristics, which include the transfer function, loaded quality factor, and effect of loss and loading, as discussed in the following section.

### A. Transfer Function and Resonance

The scattering parameter $S_{21}$ of the finite-length circuit relating the output and input is considered here as a transfer function. It is calculated using transfer matrix formalism. The reference characteristic impedance used in the definition of the scattering parameter is $R_0 = R_S = R_L = 50\,\Omega$, where $R_S$ and $R_L$ are the source and load impedances, shown in Fig. 3. Under this assumption, the $S_{21}$ parameter coincides with the voltage transfer function as

$$S_{21} = V_{out2}/V_{incident2}, \quad (10)$$

where $V_{out2}$ is the voltage at the second ladder output closed on the load $R_L$ and $V_{incident2}$ is the incident input voltage on the second ladder, that is found as $V_{incident2} = (V_{in2} + I_{in2}R_0)/2 = V_s/2$. Hence, we only excite the second ladder and we assume that $R_s = R_L = 50\,\Omega$, while all the other terminals are short circuited.

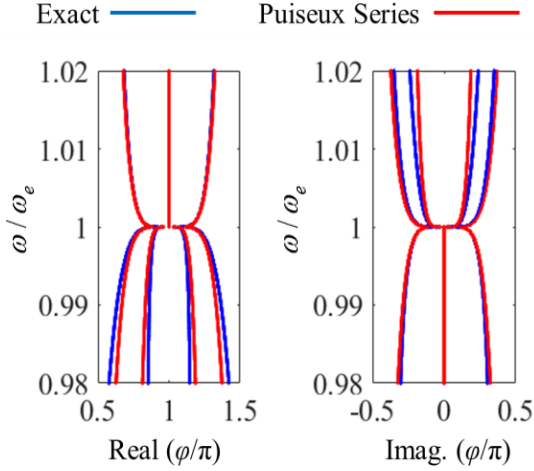

Fig. 2. Complex phase-frequency dispersion relation of the infinite periodic circuit based on the first order Puiseux series approximation (solid red) in comparison with exact numerical dispersion relation (solid blue) for the lossless circuit. The propagation phase angle $\phi$ is a complex number, both the real and imaginary parts are shown.

The magnitude of the $S_{21}$ is plotted versus normalized angular frequency around the 6DBE frequency for different numbers of unit cells ($N$) in Fig. 4(a). Results are shown for $N=8$ (dashed blue), $N=10$ (dotted black), and $N=12$ (solid red). We observe multiple resonance peaks near the 6DBE angular frequency $\omega_e$, where the sharpest and the closest to $\omega_e$ is called the 6DBE resonance denoted by $\omega_{r,e}$ and this resonance features the highest loaded quality factor ($Q_{tot}$) compared to the other resonances. As the number of unit cells ($N$) increases, the 6DBE resonance gets closer to $\omega_e$ and also gets sharper, i.e., higher quality factor.

Next, we investigate the impact of another sort of perturbation, losses in the system, especially on the $S_{21}$ parameter. The presence of losses impacts the quality factor of the lumped structure due to the high sensitivity to perturbations near the 6DBE. We now study the effect of introducing a finite quality factor of the lumped components, where for simplicity, we assume that all the elements have the same quality factor $Q_e$ due to a series resistance for each $L$ and a parallel resistance for each $C$. In Fig. 4(b), we show the magnitude of $S_{21}$ parameter of the triple ladder made of 8 unit cells, i.e., $N = 8$, for different $Q_e$, where we plot the magnitude of the voltage transfer function with and without considering the loss effects. Furthermore, as shown in Fig. 4(b), when losses are introduced into each $L$ and $C$ in the circuit, the quality factor of the resonance associated to 6DBE significantly declines, compared to other resonances of the circuit. Increasing losses beyond a certain limit deteriorate the 6DBE resonance, and the transfer function's peak will completely vanish for sufficiently low $Q_e$, as for the case with $Q_e = 500$ in Fig. 4(b). This is still a high quality factor for most of the available lumped components, but at the same time, it is important to well understand the limiting factors to realize resonances in lumped ladder circuits. Furthermore, high values of high quality factors can be realized in several technologies when using distributed elements, as in wave propagation in waveguides, and the results in this paper can be helpful to better understand the physics of cavities using waveguides with a 6DBE. Finally, it is also customary to use lumped elements to approximated realistic waveguide structures when performing time domain simulation using certain commercial software, like Candence etc., further justifying the importance of this study.

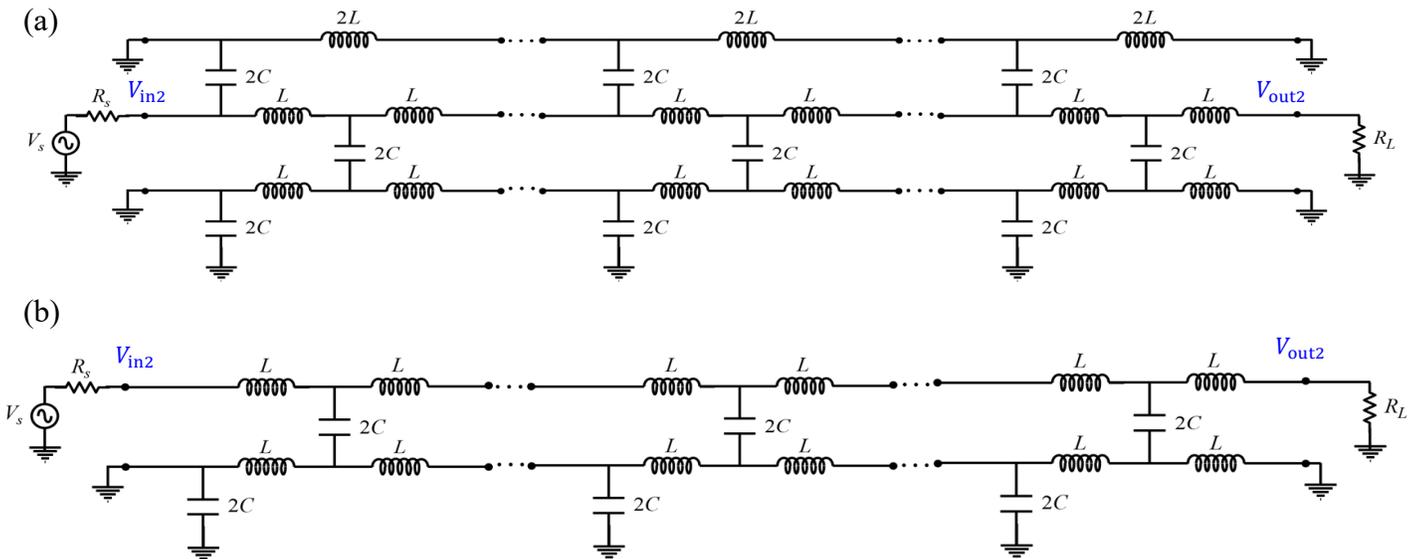

Fig. 3 (a) Triple ladder and (b) double ladder periodic *LC* circuits with finite length made of *N* unit cells with excitation voltage $V_s$, source resistance of $R_s = 50\,\Omega$, and load resistance $R_L = 50\,\Omega$ at the second (middle) ladder.

*B. Quality Factor Analysis*

As mentioned earlier, one of the main advantages of the topologies exhibiting higher orders of degeneracy (here of order six) is the enhanced quality factor compared to designs in the same technology featuring only RBE or DBE. Therefore, the theoretical investigation of the $Q$-factor is a valuable tool to analyze the performance of the 6DBE triple ladder circuit. We study the effect of losses as well as the length of the periodic triple ladder resonator circuit on the loaded quality factor $Q_{tot}$. Here the loaded $Q$-factor accounts for port terminations of 50 $\Omega$ resistive loads and losses of the reactive elements. We calculate it at the 6DBE resonance frequency $\omega_{r,e}$, and it is defined as [27]

$$Q_{tot} = \omega_{r,e} \frac{W_e + W_m}{P_l}, \quad (11)$$

in which $P_l$ is the power loss defined as the time-average dissipated power in all elements as well as in resistive loads, $W_e$ and $W_m$ are the total time-average stored energies in the capacitors and inductors of the resonator circuit, respectively, and $\omega_{r,e}$ is the resonance angular frequency closest to the 6DBE resonance. A thorough study of the loaded quality factor for a double ladder circuit featuring the DBE for the lossless and lossy cases was provided in Ref. [16]. Here, we consider the loaded $Q$-factor of the triple ladder circuit exhibiting 6DBE for both lossless and lossy cases and compare the results to that of the DBE structures. First, we explore the limiting effect of losses on the $Q$-factor. Losses are represented by finite quality factor $Q_e$ for each $L$ with a series resistance and for each $C$ with a parallel resistance.

The $Q$-factor results of both Fig. 5 plots are calculated for the case where we only excite the second ladder of the 6DBE triple ladder circuit, with the assumption of $R_s = R_L = 50$ $\Omega$ and short circuit for all the other ports as shown in Fig. 3 (b). For the DBE circuit results, the excitation is at the top line with the assumption of $R_s = R_L = 50$ $\Omega$ and short circuit for the bottom line's terminations as shown in Fig. 3(a).

In Fig. 5(a), the total loaded $Q$-factor, $Q_{tot}$, is plotted versus the elements' quality factor, $Q_e$, for different circuits lengths $N$ where we assume the same $Q_e$ for all the inductors and capacitors. By increasing $Q_e$, $Q_{tot}$ initially increases linearly, and when increasing $Q_e$ after a certain level (depending on the size of the circuit) the curves flatten out because of the effect of the 50 $\Omega$ resistive loads at the terminals. While as discussed in [15], the $Q_{tot}$ has a slight dependency on the load terminations for the resonance associated to the degeneracy condition, by using very low impedance (e.g. short circuit) or very high impedance (e.g. open circuit) terminating loads, the $Q_{tot}$ of the circuit might be enhanced, though this approach will lessen the amount of power available at the load terminals which is the case for our 6DBE triple-ladder circuit as well. Hence, here in our analysis and simulations we are assuming resistive load values of 50 $\Omega$ as terminations.

Then, we want to gain a better understanding of how the total $Q$-factor of the 6DBE circuit is affected by changing the length $N$ and loss of the elements in comparison to the DBE circuit (which has a 4$^{th}$ order degeneracy). In Fig. 5(b), we have plotted the $Q_{tot}$ versus the number of unit cells $N$ in the periodic circuit for both the 6DBE triple-ladder and the DBE double-ladder resonator circuits for lossless, as well as for two different lossy cases, with $Q_e = 10^4$ and $Q_e = 10^5$.

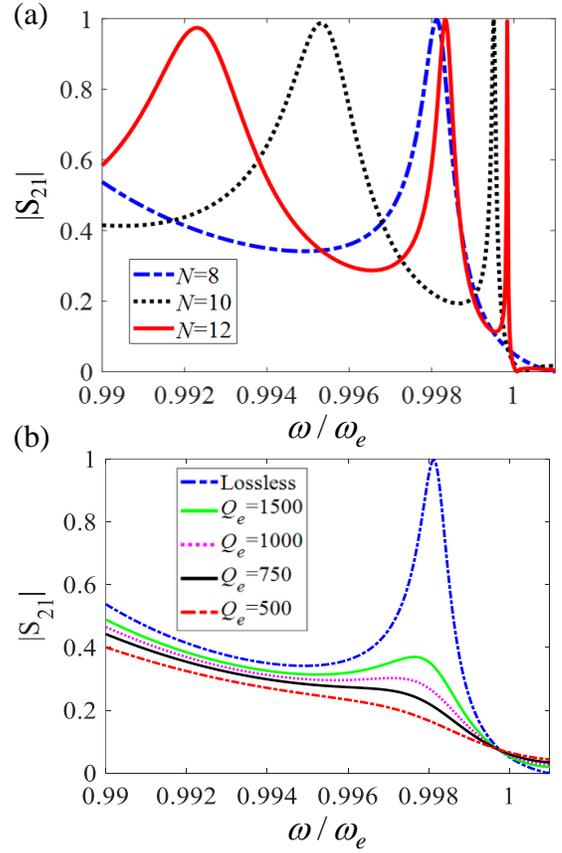

Fig. 4. Scattering parameter $S_{21}$ of the circuit shown in Fig. 3 versus frequency for (a) different number of unit cells and no loss in the elements, and (b) different quality factors of the elements in an $N=8$ unit-cells resonator. The important resonance is the one closest to $\omega_e$, denoted as $\omega_{r,e}$.

As shown in [15] for the lossless double ladder circuit with DBE, $Q_{tot}$ is asymptotically proportional to $N^5$ for large enough lengths $N$. This was also the case for other structures exhibiting DBE as shown in [10], [28], [29]. Instead, our analysis of the triple ladder 6DBE circuit demonstrates the proportionality of $N^7$ for $Q_{tot}$, for large numbers of unit cells in the lossless circuit. Consequently, the advantage of 6DBE scheme is better observed for larger ladder lengths as shown in Fig. 5(b) where, for example, for $N = 12$ unit cells, the $Q_{tot}$ of the lossless 6DBE design is almost one order of magnitude higher than that of the DBE scheme providing a substantial improvement. However, for the cases with lossy reactive elements, the loaded quality factor does not grow indefinitely with $N^7$ and it is rather limited by the finite quality factor of the elements as shown in Fig. 5(b) for $Q_e = 10^4$ and $Q_e = 10^5$. Therefore, as observed from the results of our analysis, the quality factor for the lossy cases grows exponentially only for small values of $N$, then declines and becomes saturated for larger values of $N$. For $Q_e = 10^4$, $Q_{tot}$ shows little variation with $N$ since it flattens also at $N = 8$. However, the advantage of the 6DBE over the 4th order DBE schemes in terms of higher $Q$-factor is observed even when loss is present, i.e., for $Q_e = 10^5$.

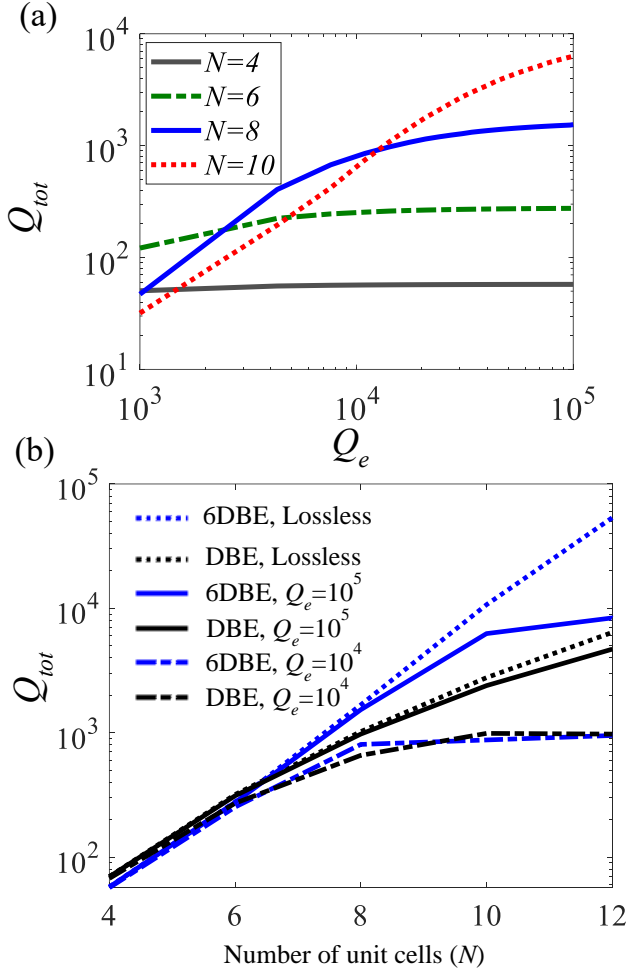

Fig. 5. (a) Loaded quality factor, $Q_{tot}$, versus elements' quality factor $Q_e$, for different numbers of unit cells $N$. (b) Loaded quality factor, $Q_{tot}$, versus number of unit cells, $N$, for different elements' quality factors. The total quality factor of the 6DBE resonator is higher than that of the DBE resonator for $Q_e$ equal or larger than $10^5$.

## IV. CONCLUSION

We have presented a resonator concept made by cascading unit cells of a triple-ladder circuit. This leads to a strong degeneracy of order six (6DBE) in its phase-frequency dispersion relation and in the state vectors. The proposed design exploits the features and enhancements related to degeneracies of order higher than those previously introduced with double ladder (exhibiting DBE) and single ladder (exhibiting only RBE) circuits, particularly in terms of $Q$-factor and sensitivity to perturbations. The approximation of the dispersion diagram obtained when using the Puiseux series demonstrates a remarkable agreement with the dispersion obtained numerically, for frequencies close to the 6DBE one.

The DBE concept has been already demonstrated useful to conceive DBE oscillators at radio frequency (RF) using lumped elements [16], [30] microstrips [22], and at optical frequency [31]. Recently the DBE oscillators have been experimentally demonstrated in microwave technology [32].

A 6DBE theoretical study showed a convenience in obtaining low threshold lasers [23], and we expect analogous advantages in using the 6DBE for oscillators at RF frequencies.

Studies of degeneracies in the triple ladder of lumped elements are useful also for understanding the behavior of distributed triple ladder waveguides, that offer higher quality factors than circuits with lumped elements; this includes also the capability to implement time domain simulations of waveguide structures in circuit simulators that are often done by approximating the circuit with lumped elements.

The analytical framework established here may be applied to a variety of structure designs and applications featuring degeneracy conditions specifically, but not limited to, sixth order and will be a useful tool in designing and evaluating novel designs for high $Q$-factor, oscillators, pulse generators, sensory and filter applications.

ACKNOWLEDGMENT

This material is based upon work supported by the National Science Foundation, USA, under award NSF ECCS-1711975.

**Appendix: Transfer Matrix Formalism**

The transfer matrix $\underline{\mathbf{T}}$ of the unit cell of the triple ladder circuit in Fig. 6 is obtained by multiplying the $6 \times 6$ transfer matrices of each segment, calculated based on the state vector definition, in the proper order as

$$\underline{\mathbf{T}} = \underline{\mathbf{T}}_4 \underline{\mathbf{T}}_3 \underline{\mathbf{T}}_2 \underline{\mathbf{T}}_1.$$

The calculation of $\underline{\mathbf{T}}$ is burdensome and after simplifications it results in the following matrix:

$$\underline{\mathbf{T}} = \begin{bmatrix} 1-4a & 4a & 0 & -Z_L & 0 & 0 \\ 4a(1-a) & a(4a-6)+1 & 2a(1-2a) & 0 & Z_L(a-1) & -Z_L a \\ 4a^2 & 2a(1-2a) & a(4a-6)+1 & 0 & -Z_L a & Z_L(a-1) \\ -Y_C & Y_C & 0 & 1 & 0 & 0 \\ Y_C(1-2a) & 2Y_C(a-1) & Y_C(1-2a) & 0 & 1-2a & 2a \\ 2Y_C a & Y_C(1-2a) & 2Y_C(a-1) & 0 & 2a & 1-2a \end{bmatrix},$$

where $a = LC\omega^2 = (\omega/\omega_e)^2$, $Z_L = 2j\omega L$, and $Y_C = 2j\omega C$.

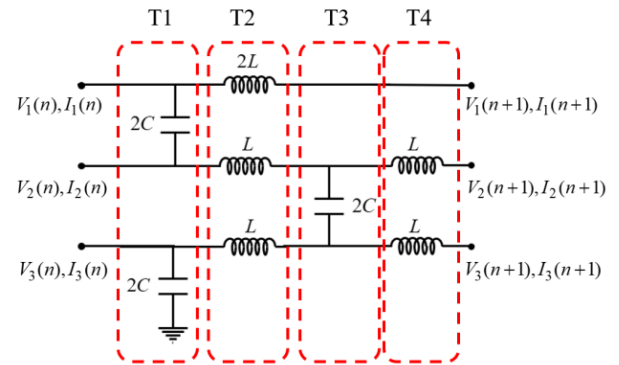

Fig. 6. Segments of the unit cell of a periodic triple-ladder lumped circuit to find transfer matrix $\underline{\mathbf{T}}$.

Using the formulation described above, the transfer matrix of the one unit cell of the lossless circuit evaluated exactly at 6DBE frequency is calculated to be

$$\underline{\mathbf{T}}_e = \begin{bmatrix} -3 & 4 & 0 & -2j\sqrt{\frac{L}{C}} & 0 & 0 \\ 0 & -1 & -2 & 0 & 0 & -2j\sqrt{\frac{L}{C}} \\ 4 & -2 & -1 & 0 & -2j\sqrt{\frac{L}{C}} & 0 \\ -2j\sqrt{\frac{C}{L}} & 2j\sqrt{\frac{C}{L}} & 0 & 1 & 0 & 0 \\ -2j\sqrt{\frac{C}{L}} & 0 & -2j\sqrt{\frac{C}{L}} & 0 & -1 & 2 \\ 4j\sqrt{\frac{C}{L}} & -2j\sqrt{\frac{C}{L}} & 0 & 0 & 2 & -1 \end{bmatrix}.$$

The 6DBE is an exceptional point of degeneracy, and the transfer matrix $\underline{\mathbf{T}}_e$ is similar to a matrix with a Jordan canonical form as $\underline{\mathbf{T}}_e = \underline{\mathbf{V}}\underline{\mathbf{\Lambda}}_J\underline{\mathbf{V}}^{-1}$ where $\underline{\mathbf{\Lambda}}_J$ has the form of

$$\underline{\mathbf{\Lambda}}_J = \begin{pmatrix} -1 & 1 & 0 & 0 & 0 & 0 \\ 0 & -1 & 1 & 0 & 0 & 0 \\ 0 & 0 & -1 & 1 & 0 & 0 \\ 0 & 0 & 0 & -1 & 1 & 0 \\ 0 & 0 & 0 & 0 & -1 & 1 \\ 0 & 0 & 0 & 0 & 0 & -1 \end{pmatrix},$$

when the degenerate eigenvalue $\lambda = \exp(-j\varphi)$ is evaluated at $\varphi = \pi$.

## V. REFERENCES

[1] N. Dunford, J. T. Schwartz, W. G. Bade, and R. G. Bartle, *Linear operators*. New York: Wiley interscisense, 1958.

[2] T. Kato, *Perturbation theory for linear operators*. Verlag, NY: Springer Science & Business Media, 2013.

[3] A. Figotin and I. Vitebskiy, "Slow light in photonic crystals," *Waves Random Complex Media*, vol. 16, no. 3, pp. 293–382, 2006.

[4] A. Figotin and I. Vitebskiy, "Slow wave phenomena in photonic crystals," *Laser Photonics Rev.*, vol. 5, no. 2, pp. 201–213, 2011.

[5] M. A. Othman and F. Capolino, "Demonstration of a degenerate band edge in periodically-loaded circular waveguides," *IEEE Microw. Wirel. Compon. Lett.*, vol. 25, no. 11, pp. 700–702, 2015.

[6] M. A. K. Othman, F. Yazdi, A. Figotin, and F. Capolino, "Giant gain enhancement in photonic crystals with a degenerate band edge," *Phys. Rev. B*, vol. 93, no. 2, p. 024301, Jan. 2016, doi: 10.1103/PhysRevB.93.024301.

[7] A. F. Abdelshafy, M. A. K. Othman, D. Oshmarin, A. T. Almutawa, and F. Capolino, "Exceptional Points of Degeneracy in Periodic Coupled Waveguides and the Interplay of Gain and Radiation Loss: Theoretical and Experimental Demonstration," *IEEE Trans. Antennas Propag.*, vol. 67, no. 11, pp. 6909–6923, Nov. 2019, doi: 10.1109/TAP.2019.2922778.

[8] C. Löcker, K. Sertel, and J. L. Volakis, "Emulation of propagation in layered anisotropic media with equivalent coupled microstrip lines," *Microw. Wirel. Compon. Lett. IEEE*, vol. 16, no. 12, pp. 642–644, 2006.

[9] J. L. Volakis and K. Sertel, "Narrowband and Wideband Metamaterial Antennas Based on Degenerate Band Edge and Magnetic Photonic Crystals," *Proc. IEEE*, vol. 99, no. 10, pp. 1732–1745, Oct. 2011, doi: 10.1109/JPROC.2011.2115230.

[10] M. A. K. Othman, X. Pan, G. Atmatzakis, C. G. Christodoulou, and F. Capolino, "Experimental Demonstration of Degenerate Band Edge in Metallic Periodically Loaded Circular Waveguide," *IEEE Trans. Microw. Theory Tech.*, vol. 65, no. 11, pp. 4037–4045, Nov. 2017, doi: 10.1109/TMTT.2017.2706271.

[11] N. Gutman, C. Martijn de Sterke, A. A. Sukhorukov, and L. C. Botten, "Slow and frozen light in optical waveguides with multiple gratings: Degenerate band edges and stationary inflection points," *Phys. Rev. A*, vol. 85, no. 3, Mar. 2012, doi: 10.1103/PhysRevA.85.033804.

[12] M. G. Wood, J. R. Burr, and R. M. Reano, "Degenerate band edge resonances in periodic silicon ridge waveguides," *Opt. Lett.*, vol. 40, no. 11, pp. 2493–2496, 2015.

[13] M. Y. Nada, M. A. K. Othman, and F. Capolino, "Theory of coupled resonator optical waveguides exhibiting high-order exceptional points of degeneracy," *Phys. Rev. B*, vol. 96, no. 18, p. 184304, Nov. 2017, doi: 10.1103/PhysRevB.96.184304.

[14] M. Y. Nada, M. A. K. Othman, O. Boyraz, and F. Capolino, "Giant Resonance and Anomalous Quality Factor Scaling in Degenerate Band Edge Coupled Resonator Optical Waveguides," *J. Light. Technol.*, vol. 36, no. 14, pp. 3030–3039, Jul. 2018, doi: 10.1109/JLT.2018.2822600.

[15] J. T. Sloan, M. A. K. Othman, and F. Capolino, "Theory of Double Ladder Lumped Circuits With Degenerate Band Edge," *IEEE Trans. Circuits Syst. Regul. Pap.*, vol. 65, no. 1, pp. 3–13, Jan. 2018, doi: 10.1109/TCSI.2017.2690971.

[16] D. Oshmarin *et al.*, "A New Oscillator Concept Based on Band Edge Degeneracy in Lumped Double-Ladder Circuits," *IET Circuits Devices Amp Syst.*, Feb. 2019, doi: 10.1049/iet-cds.2018.5048.

[17] J. P. Dowling, M. Scalora, M. J. Bloemer, and C. M. Bowden, "The photonic band edge laser: A new approach to gain enhancement," *J. Appl. Phys.*, vol. 75, no. 4, pp. 1896–1899, Feb. 1994, doi: 10.1063/1.356336.

[18] G. L. Matthaei, L. Young, and E. M. Jones, "Design of microwave filters, impedance-matching networks, and coupling structures. Volume 2," DTIC Document, 1963.

[19] E. A. Guillemin, *Synthesis of passive networks: theory and methods appropriate to the realization and approximation problems*. RE Krieger Pub. Co., 1977.

[20] B. Razavi, *Design of Integrated Circuits for Optical Communications*. Hoboken, NJ: Wiley, 2012.

[21] V. A. Tamma, A. Figotin, and F. Capolino, "Concept for Pulse Compression Device Using Structured Spatial

Energy Distribution," *IEEE Trans. Microw. Theory Tech.*, vol. 64, no. 3, pp. 742–755, Mar. 2016, doi: 10.1109/TMTT.2016.2518160.

[22] A. F. Abdelshafy, D. Oshmarin, M. A. K. Othman, M. M. Green, and F. Capolino, "Distributed Degenerate Band Edge Oscillator," *IEEE Trans. Antennas Propag.*, vol. 69, no. 3, pp. 1821–1824, Mar. 2021, doi: 10.1109/TAP.2020.3018539.

[23] M. Y. Nada and F. Capolino, "Exceptional point of sixth-order degeneracy in a modified coupled-resonator optical waveguide," *JOSA B*, vol. 37, no. 8, pp. 2319–2328, Aug. 2020, doi: 10.1364/JOSAB.385198.

[24] A. Welters, "On Explicit Recursive Formulas in the Spectral Perturbation Analysis of a Jordan Block," *SIAM J. Matrix Anal. Appl.*, vol. 32, no. 1, pp. 1–22, Jan. 2011, doi: 10.1137/090761215.

[25] T. Kato, *Perturbation theory for linear operators*, vol. 132. springer, 1995.

[26] F. Zhang, "Quaternions and matrices of quaternions," *Linear Algebra Its Appl.*, vol. 251, pp. 21–57, Jan. 1997, doi: 10.1016/0024-3795(95)00543-9.

[27] Pozar, David, *Microwave Engineering*. Hoboken, NJ: Wiley, 2012.

[28] M. A. K. Othman, F. Yazdi, A. Figotin, and F. Capolino, "Giant gain enhancement in photonic crystals with a degenerate band edge," *Phys. Rev. B*, vol. 93, no. 2, p. 024301, Jan. 2016.

[29] J. R. Burr, N. Gutman, C. Martijn de Sterke, I. Vitebskiy, and R. M. Reano, "Degenerate band edge resonances in coupled periodic silicon optical waveguides," *Opt. Express*, vol. 21, no. 7, pp. 8736–8745, Apr. 2013, doi: 10.1364/OE.21.008736.

[30] M. Radfar, D. Oshmarin, M. A. K. Othman, M. M. Green, and F. Capolino, "Low Phase Noise Oscillator Design Using Degenerate Band Edge Ladder Architectures," *IEEE Trans. Circuits Syst. II Express Briefs*, pp. 1–1, 2021, doi: 10.1109/TCSII.2021.3088322.

[31] M. Veysi, M. A. K. Othman, A. Figotin, and F. Capolino, "Degenerate band edge laser," *Phys. Rev. B*, vol. 97, no. 19, p. 195107, May 2018, doi: 10.1103/PhysRevB.97.195107.

[32] D. Oshmarin, A. F. Abdelshafy, A. Nikzamir, and M. M. Green, and F. Capolino, "Experimental Demonstration of a New Oscillator Concept Based on Degenerate Band Edge in Microstrip Circuit," *arXiv preprint arXiv:2109.07002*, 2021.